\documentclass[amsmath,amssymb,11pt,superscriptaddress,reprint, preprintnumbers, notitlepage,aps,prl,twocolumn]{revtex4-1}
\pdfoutput=1 
\usepackage[utf8]{inputenc}
\usepackage[english]{babel}
\usepackage{amsmath}
\usepackage{graphicx}
\usepackage{dcolumn}
\usepackage{pbox}
\usepackage{amssymb}
\usepackage{epsfig}
\usepackage[dvipsnames]{xcolor}
\usepackage{slashed}
\usepackage{amssymb}
\usepackage{ mathrsfs }
\usepackage{color}
\usepackage[font=small]{caption}
\usepackage[font=small]{subcaption}
\usepackage{url}
\definecolor{MyDarkBlue}{rgb}{0.1, 0.1, 0.8} 
\definecolor{SBlue}{rgb}{0.2, 0.4, 0.7} 
\definecolor{MyLightBlue}{rgb}{0.22,0.51,0.9}
\definecolor{MyGreen}{rgb}{0.0, 0.5, 0.0}
\definecolor{BrickRed}{rgb}{0.8, 0.25, 0.33}
\RequirePackage{hyperref}
\hypersetup{colorlinks, citecolor=SBlue,linkcolor=MyDarkBlue, urlcolor=PineGreen}

\newcommand{\sigmav}{\ensuremath{{\langle \sigma v \rangle}}}

\newcommand{\MeV}{\,\mathrm{MeV}}
\newcommand{\GeV}{\,\mathrm{GeV}}

\makeatletter

\makeatletter
\renewcommand\@makecaption[2]{%
  \par
  \vskip\abovecaptionskip
  \begingroup
  
   \small\rmfamily
    \begingroup
     \samepage
     \flushing
     \let\footnote\@footnotemark@gobble
     \@make@capt@title{#1}{#2}\par
    \endgroup
  \endgroup
  \vskip\belowcaptionskip
}
\makeatother

\DeclareUnicodeCharacter{2212}{-}
\begin{document}
\title{\vspace{1cm}\Large 
Minimal realization of light thermal Dark Matter
}

\author{\bf Johannes Herms}
\email[E-mail:]{herms@mpi-hd.mpg.de}
\affiliation{Max-Planck-Institut f{\"u}r Kernphysik, Saupfercheckweg 1, 69117 Heidelberg, Germany}
\author{\bf Sudip Jana}
\email[E-mail:]{sudip.jana@mpi-hd.mpg.de}
\affiliation{Max-Planck-Institut f{\"u}r Kernphysik, Saupfercheckweg 1, 69117 Heidelberg, Germany}
\author{\bf Vishnu P.K.}
\email[E-mail:]{ vipadma@okstate.edu}
\affiliation{Department of Physics, Oklahoma State University, Stillwater, OK, 74078, USA}
\author{\bf Shaikh Saad}
\email[E-mail:]{ shaikh.saad@unibas.ch}
\affiliation{Department of Physics, University of Basel, Klingelbergstrasse 82, CH-4056 Basel, Switzerland}

\begin{abstract}
We propose a minimal UV-complete model for kinematically forbidden Dark Matter (DM) leading to a sub-GeV thermal relic. Our crucial realization is that the two-Higgs-doublet model can provide a light mediator through which the DM can annihilate into SM leptons, avoiding indirect detection constraints. The DM mass is predicted to be very close to the mass of the leptons, which can potentially be identified from DM annihilation into gamma-rays. Due to sizable couplings to muons in reproducing the DM relic abundance, this framework naturally favors a resolution to the $(g-2)_\mu$ anomaly. Furthermore, by embedding this setup to the Zee model, we show that the phenomenon of neutrino oscillations is inherently connected to the observed relic abundance of DM. All new physics involved in our framework lies at or below the electroweak scale, making it testable at upcoming colliders, beam-dump experiments, and future sub-GeV gamma-ray telescopes.
\end{abstract}

\maketitle
\textbf{\emph{Introduction}.--} 
Dark matter (DM) is a central part of our understanding of the cosmos, and identifying its nature is a key goal of contemporary cosmology, astro- and particle physics.
A prime candidate for DM is thermal relic particles -- a new neutral, long-lived particle species that was in thermal equilibrium with the particles of the Standard Model (SM) in the early Universe before the connecting interactions froze-out when it reached the DM abundance we observe today.

This archetypal Weakly Interacting Massive Particle (WIMP) DM scenario motivated great experimental efforts to identify these particles. Traditionally, the focus has been on electroweak-scale WIMPs (see eg.~\cite{Bertone:2016nfn,Lee:1977ua}) since these
are expected in many theories beyond the Standard Model and can naturally decouple at the correct abundance.
With, on the one hand, the experimental program to search for EW-scale DM particles well underway, and on the other hand, the degree of confidence in new physics at the EW-scale waning in the face of the success of the SM at the LHC, thermal relics at smaller masses have become a focus of attention. 
This trend in DM studies is supported by connections to low-energy anomalies in particle physics, in particular, the $(g-2)_\mu$ tension \cite{Muong-2:2021ojo}, and can relate to neutrino physics, see, e.g., Ref.~\cite{Jana:2020joi}.

\begin{figure}[b!]
$$
\includegraphics[width=0.37\textwidth]{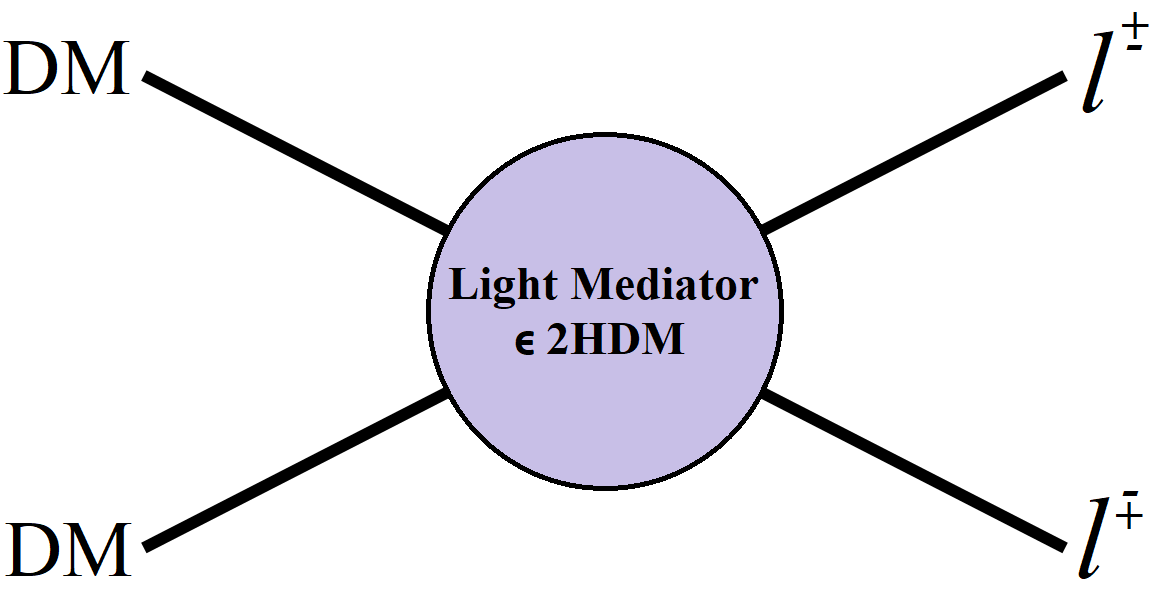}
$$
\caption{Schematic diagram for light DM annihilation to SM charged leptons via 2HDM-portal.
} \label{schematic}
\end{figure}

Thermal DM in the mass range $\MeV \lesssim m_\mathrm{DM} \lesssim 10 \GeV$ requires a few general conditions.
It typically requires a stabilizing symmetry to forbid decay, in contrast to lighter, keV-scale DM candidates like sterile neutrinos~\cite{Dodelson:1993je,Roach:2019ctw,Heeba:2018wtf}. Equilibration with the SM in the early Universe then proceeds via annihilation, as in the WIMP scenario.
The annihilation products cannot be much heavier than the DM candidate, necessitating a coupling to light SM particles.
In contrast to heavier WIMPs, successful sub-GeV freeze-out DM generically requires the existence of a new light mediator to enable a sufficiently large annihilation rate \footnote{This is true also in scenarios where DM does not annihilate directly to SM particles, such as secluded DM~\cite{Pospelov:2007mp} or to a lesser extent SIMP DM~\cite{Hochberg:2015vrg}, where the heat produced in DM annihilation is ultimately transferred to the SM bath}. For this reason, many models in the literature propose to extend the gauge symmetry \footnote{For non-Abelian gauge extension that offers vector dark matter candidate, see, e.g., Ref.  \cite{Chowdhury:2021tnm}.} of the theory by a $U(1)_D$, with a dark photon that can mediate between the dark and visible sectors (for instance, see e.g., Refs.~\cite{Boehm:2003hm,DAgnolo:2015ujb,Lee:2015gsa,Pospelov:2007mp}).

In contrast to these previous DM theories that include new gauge sectors (e.g.~\cite{Boehm:2003hm,DAgnolo:2015ujb,Lee:2015gsa,Pospelov:2007mp}) or multi-fermionic extensions (e.g.~\cite{Boehm:2003hm,Batell:2017kty,DAgnolo:2020mpt}), this letter presents a minimal, ultraviolet complete (UV) model of sub-GeV thermal DM. Our proposal is based on the crucial fact that adding a second Higgs doublet \cite{Branco:2011iw} to the SM allows for a light scalar \cite{Jana:2020pxx} that can couple to the light SM degrees of freedom and DM particles. Specifically, we work in the two-Higgs-doublet model (2HDM) framework, where DM annihilates into SM leptons via a light mediator emerging from the 2HDM.  The simplest DM candidate is a real scalar stabilized by a $\mathbb{Z}_2$ symmetry that can easily reproduce the observed relic abundance utilizing the 2HDM-portal.

After taking into account constraints on the DM annihilation today and during the epoch of CMB decoupling, the DM mass is required to be in the `forbidden' regime~\cite{Griest:1990kh,DAgnolo:2015ujb,DAgnolo:2020mpt}, where it is just slightly lighter than the SM particle it annihilates into. This makes for a predictive and positively identifiable framework, motivating particularly for searches for sub-GeV gamma-ray lines close to SM lepton masses in the simplest scenario. 
Furthermore, the favored parameter space implies a positive shift for muon $g-2$ with the proper sign  and strength to account for the measurement at Fermilab \cite{Muong-2:2021ojo}.

Additionally, we consider the possibility of neutrino mass generation by a simple extension of this setup. Introducing a charged scalar allows for non-zero neutrino mass via one-loop quantum corrections. Within this framework, the same Yukawa couplings reproducing the correct DM relic abundance also participate in neutrino mass generation while remaining consistent with lepton flavor violating constraints.

To demonstrate the versatility of the light 2HDM scalar portal, next, we entertain the scenario of fermionic DM (see also~\cite{Batell:2016ove,Jia:2021mwk}). A singlet fermion (either Dirac or Majorana) DM can annihilate into SM leptons via a singlet scalar that mixes with the light scalar arising from the 2HDM. Annihilation of fermion DM through a scalar mediator is velocity-suppressed, evading bounds on energy release during CMB decoupling. This allows for a wide range of DM and mediator masses, illustrating the broad potential of the light 2HDM-portal. 

In the following, before presenting minimal models for scalar and fermionic sub-GeV DM, we first recapitulate the thermal freeze-out mechanism for DM production.
We present constraints from DM relic density, $(g-2)_\mu$, flavor violation, and DM indirect detection. Before concluding, we briefly demonstrate how neutrino masses can be naturally incorporated into the light-2HDM scenario.

\textbf{\emph{Forbidden DM}.--} 
The relic density of thermal DM is calculated by tracing its evolution in the early Universe. If the decay of DM into SM particles is precluded by symmetry, the leading number-changing term in the Boltzmann equation stems from annihilations of pairs of DM particles
\begin{equation}
    \frac{d n_\mathrm{DM}}{d t} + 3 H n_\chi = \zeta \sigmav \left( n_\mathrm{DM}^2 -n_\mathrm{DM}^{\mathrm{eq}2}\right) \,,
    \label{eq:boltzmannEqn}
\end{equation}
where $n_\mathrm{DM}$ is the total DM number density, $H$ is the Hubble expansion rate, and $\sigmav$ is the thermally averaged cross-section of DM annihilating into other bath particles ($\zeta = (1/2)\, 1$ for (non-)self-conjugate DM particles).
In the freeze-out scenario, the relic density is determined by the time (marked by the corresponding temperature of the SM bath, $T_\mathrm{fo}$) when the annihilation rate drops below the Hubble rate
\begin{equation}
    \zeta \sigmav n_\mathrm{DM}^\mathrm{eq} = H \,.
    \label{eq:freezeOutCondition}
\end{equation}
It is often sufficient to work in the instantaneous freeze-out approximation, where the comoving DM density (denoted in terms of the abundance $Y\equiv n/s$, with $s$ the SM entropy density) stays constant after freeze-out
\begin{equation}
    Y_\mathrm{DM}^\mathrm{today} = Y_\mathrm{DM}^\mathrm{eq}\left(T_\mathrm{fo}\right) \,.
    \label{eq:instaFoApprox}
\end{equation}
This is to be compared to value corresponding to the observed DM density $\Omega_\mathrm{DM}h^2 = 0.12$~\cite{Planck:2018vyg},
\begin{equation}
    Y_\mathrm{DM}^\mathrm{obs} = \frac{n_\mathrm{DM}}{s_0} = \Omega_\mathrm{DM} \frac{\rho_C}{m_\mathrm{DM} s_0}
    = 4.35 \cdot 10^{-10} \left(\frac{m_\mathrm{DM}}{\GeV}\right)^{-1}\,.
    \label{eq:Yobserved}
\end{equation}
Requiring freeze-out at the correct temperature to reproduce the observed relic abundance leads to a constraint on $\sigmav$.

The generic implication of the WIMP scenario is DM annihilation with $\sigmav_\mathrm{fo} \sim \,\mathrm{pb} \,c$ at $T_\mathrm{fo}\sim m_\mathrm{DM}/20$.
This motivates the WIMP indirect detection program, looking for DM annihilation at overdense regions of the Universe today.
For sub-GeV DM however, this can be a problem, since $\sigmav_\mathrm{CMB\ epoch} = \sigmav_\mathrm{fo}$ results in excessive energy injection into the SM plasma during CMB decoupling for $m_\mathrm{DM} \lesssim 10\GeV$~\cite{Planck:2018vyg,Slatyer:2015jla}.

In the models considered in this work, tree-level annihilation channels are closed at low temperatures.
In this ``forbidden DM'' scenario~\cite{DAgnolo:2015ujb,DAgnolo:2020mpt}, DM particles $\chi\chi$ are slightly lighter than the bath particles $l_1 \bar l_2$ they annihilate into.
The leading annihilation rate $\chi\chi \to l_1 \bar l_2$ is then Boltzmann suppressed \footnote{
In the models we consider, DM freeze-out can also be determined by forbidden annihilations into two mediator particles, provided the mediator remains in equilibrium with the SM bath. We have verified that this is ensured by $\phi \leftrightarrow \nu \bar \nu ,\gamma\gamma$ processes for the relevant values of mediator-lepton couplings.
},
$\sigmav_{\chi\chi \to l_1 \bar l_2} \simeq \frac{1}{2 \zeta} \sigmav_{l_1 \bar l_2 \to \chi\chi} e^{-2 \Delta x}$
where $n_{l_i}$ denote the number density of $l_i$ particles plus antiparticles and $n_\mathrm{DM}$ denotes the total DM number density, while $\Delta \equiv (m_{l_1}+m_{l_2} -2 m_\mathrm{DM})/2 m_\mathrm{DM}$ and $x= m_\mathrm{DM}/T$.
Tree-level DM annihilation is strongly suppressed at low temperature for $\Delta>0$, evading CMB and cosmic ray probes.
Radiative annihilations, however, are not forbidden, and $\gamma$-ray line signals at energies just below SM particle masses can be a powerful and specific probe of the present scenario, as discussed later in the text.

\textbf{\emph{Models}.--} Here we present a minimal scenario for forbidden DM.   Our UV-complete model is a simple extension of the 2HDM \cite{Branco:2011iw} by an $SU(2)_L$ scalar singlet $S$, which  qualifies as a promising forbidden DM candidate.
The stability of the DM is ensured by a discrete  $\mathcal{Z}_2$ symmetry, under which only the DM transforms non-trivially.
In the Higgs basis, when only one neutral Higgs acquires a nonzero vacuum expectation value, the Higgs doublets can be parameterized as, 
\begin{align} 
	& H_1=\begin{pmatrix}
		G^{+}  \\
		\frac{1}{\sqrt{2}}(v+\phi_1^0+iG^0)    
	\end{pmatrix},   
\quad H_2=\begin{pmatrix}
		H^{+}  \\
		\frac{1}{\sqrt{2}}(\phi_2^0+iA)     
	\end{pmatrix}. 
\label{para}
\end{align}
Here $G^+$ and $G^0$ are the unphysical Goldstone modes, whereas $H^+$ and $\{\phi_1^0,\phi_2^0, A \}$ are the physical Higgs
bosons. The VEV of $H_1$, in our notation of $v\simeq 246$ GeV, governs the EW symmetry breaking.

We choose to work in the alignment limit~\cite{Branco:2011iw,Babu:2018uik,Bernon:2015qea,BhupalDev:2014bir}, where the SM Higgs $\phi_1^0\approx h$ decouples from the new  CP-even Higgs ($\phi_2^0\approx H$). The masses of physical scalar states in this limit are expressed as follows:
\begin{align}
&m^2_h= \lambda_1v^2,\;
m^2_H=\mu^2_{22}+\frac{v^2}{2}(\lambda_3+\lambda_4+\lambda_5),
\\
&m^2_A=m^2_H-v^2 \lambda_5,\;
m^2_{H^\pm}=m^2_H-\frac{v^2}{2}(\lambda_4+\lambda_5),
\\
&m_S^2=\mu_S^2+\frac{\kappa_1}{2} v^2,
\end{align}
where $\lambda_i$, $\kappa_i$, and $\mu_{22,S}$ are  parameters in the scalar potential (see Appendix-I).
Here $m_S$ denotes the mass of the DM candidate $S$. From the above mass relations, it is straightforward to see that the BSM CP-even state $H$ can be made light without making the CP-odd state $A$ and the charged
$H^\pm$ light. The emergence of this light state from the 2HDM is the key to realizing light thermal DM.
However, having a light $H$ puts an upper bound on the masses of $A$ and $H^\pm$ from EW precision constraints which are discussed in detail later.

In our analysis, both for simplicity and possible connections to muon and neutrino properties, we consider the second doublet to be leptophilic in nature. The Yukawa part of the Lagrangian then reads, 
\begin{align}
	-&\mathcal{L}_Y\supset  \widetilde{Y}_l \bar{\psi}_L H_1 \psi_R + Y_l \bar{\psi}_L H_2 \psi_R + h.c. \label{Yukawa}
\end{align}
In the alignment limit that we are working, the former Yukawa coupling is responsible for generating masses of the charged leptons, i.e., $\widetilde{Y}_l = \mathrm{diag}(m_e,m_\mu,m_\tau)/v$, while $Y_l$ determines the DM phenomenology. 
We focus on the minimal scalar DM scenario, where the structure of $Y_l$ needs to ensure only kinematically forbidden channels receive significant couplings.
In particular, forbidden DM annihilating to $\mu\mu$, $\mu\tau$ and $\tau\tau$ is phenomenologically  possible. The forbidden $ee$ channel is precluded by thermal DM of $m_\mathrm{DM}\lesssim m_e$ being too light~\cite{Boehm:2013jpa}, while $e\mu$ or $e\tau$ are constrained by lepton flavour violation~\cite{DAgnolo:2020mpt}.

We also consider fermionic forbidden DM based on the light 2HDM mediated scenario. In this case, the scalar mediator needs to have an admixture of an additional light singlet scalar to couple to a pair of DM fermions. The full model description is given in the Appendix-II. In the fermionic case, annihilations are velocity-suppressed, and non-forbidden annihilation is no longer precluded by CMB bounds, allowing in general for a wider range of DM masses and Yukawa structures $Y_l$. For the purposes of this letter, however, we restrict ourselves to forbidden mass spectra.


\begin{figure*}[t!]
$$
\includegraphics[width=0.32\textwidth]{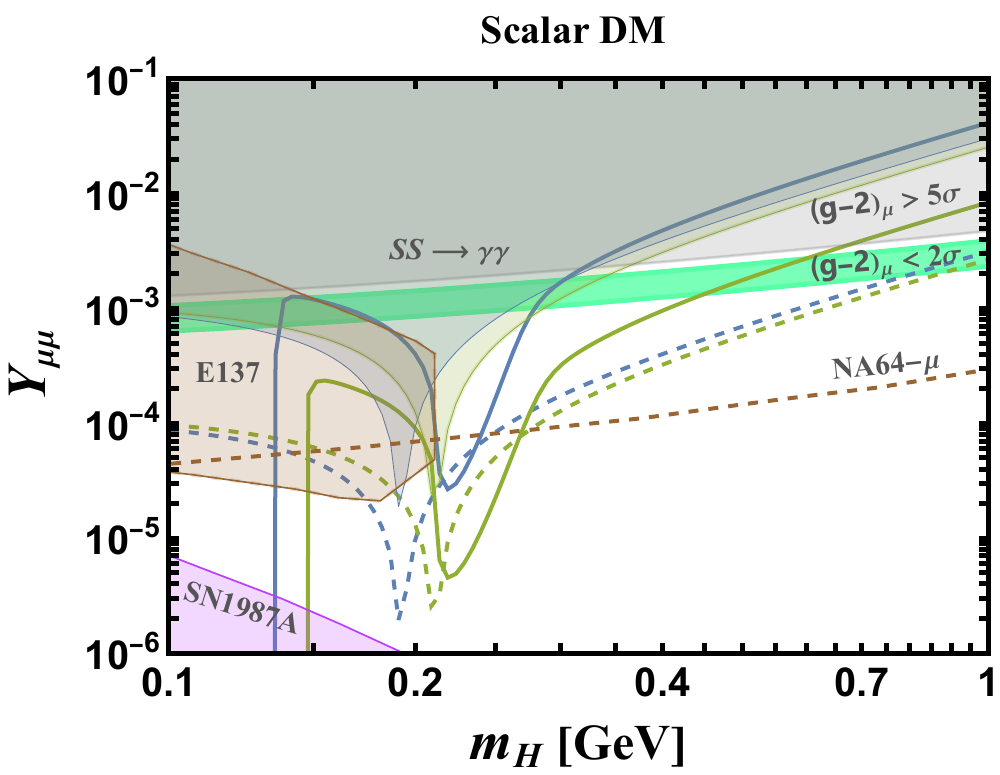}\hspace{0.12in}
\includegraphics[width=0.32\textwidth]{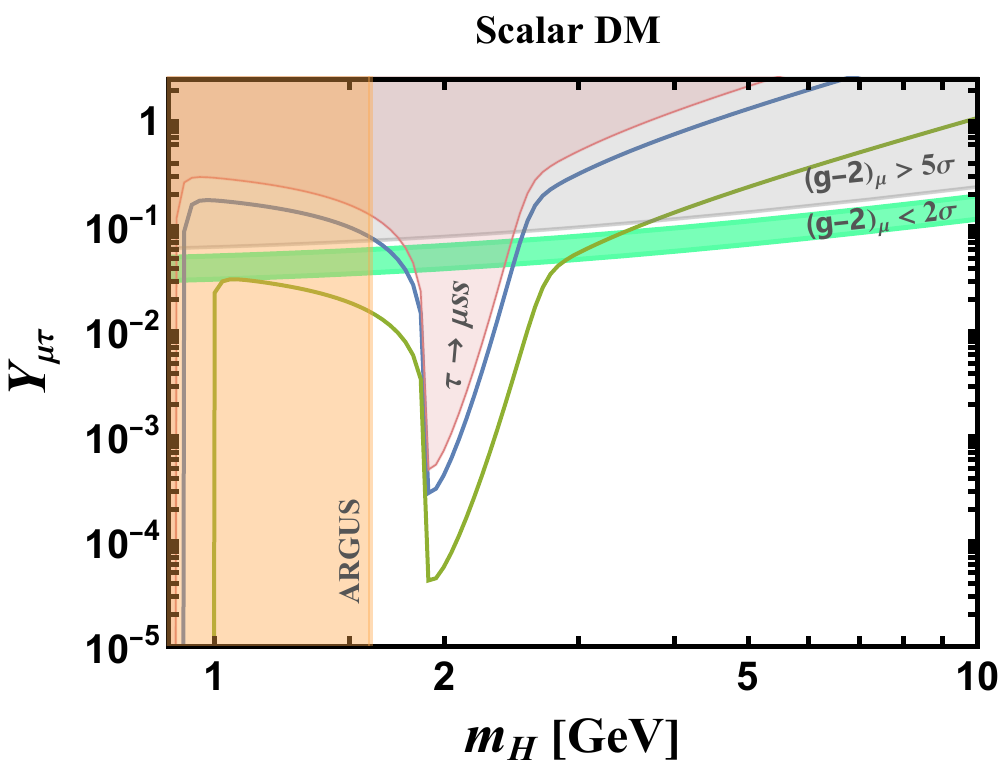}\hspace{0.12in}
\includegraphics[width=0.32\textwidth]{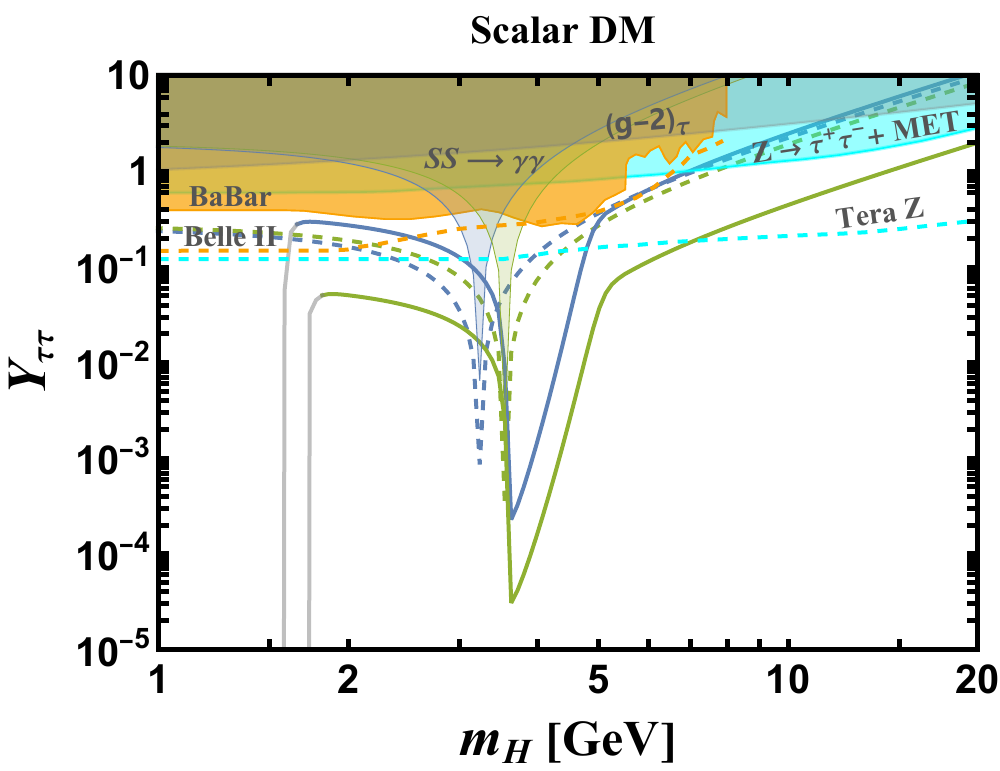}
$$
$$
\includegraphics[width=0.32\textwidth]{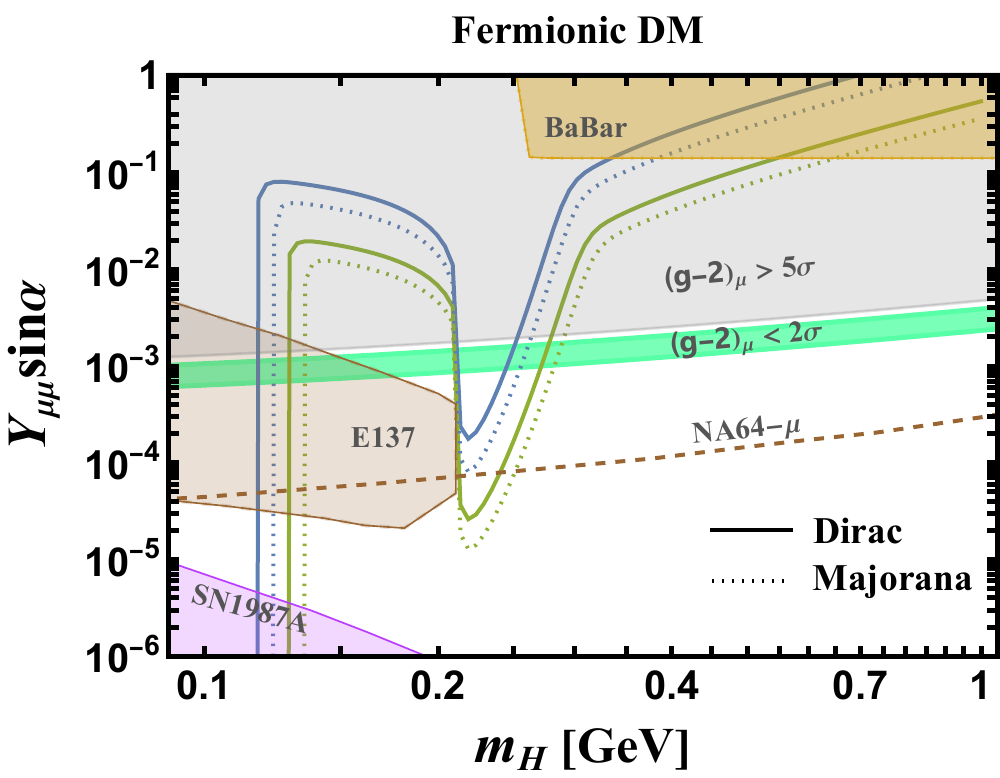}\hspace{0.12in}
\includegraphics[width=0.325\textwidth]{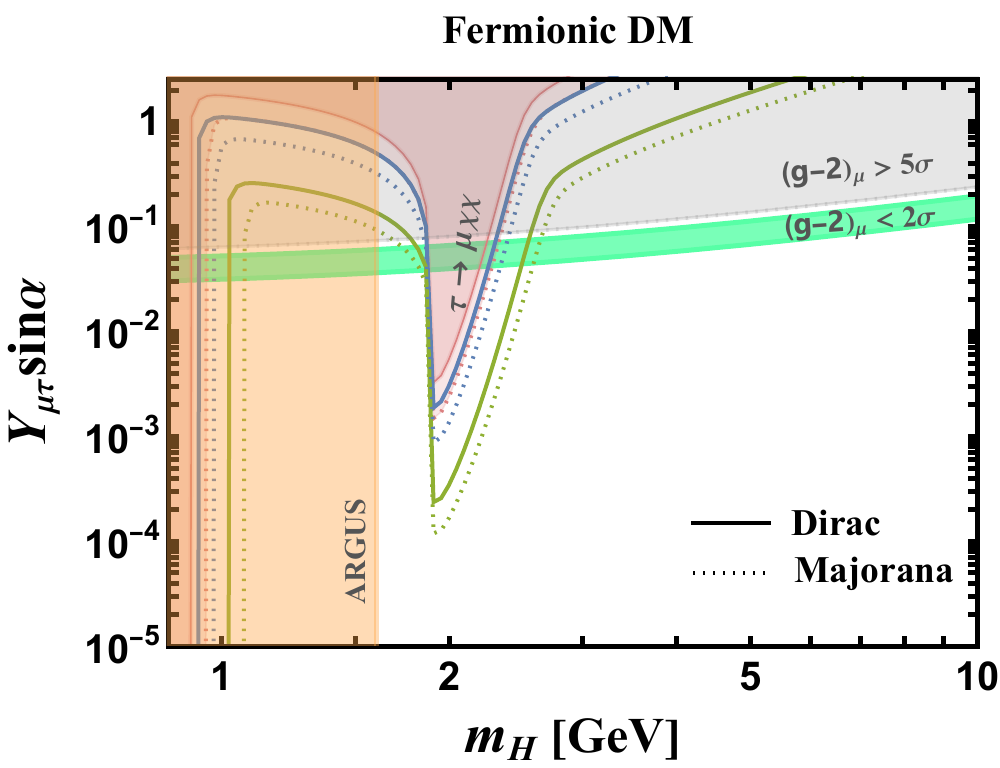}\hspace{0.12in}
\includegraphics[width=0.32\textwidth]{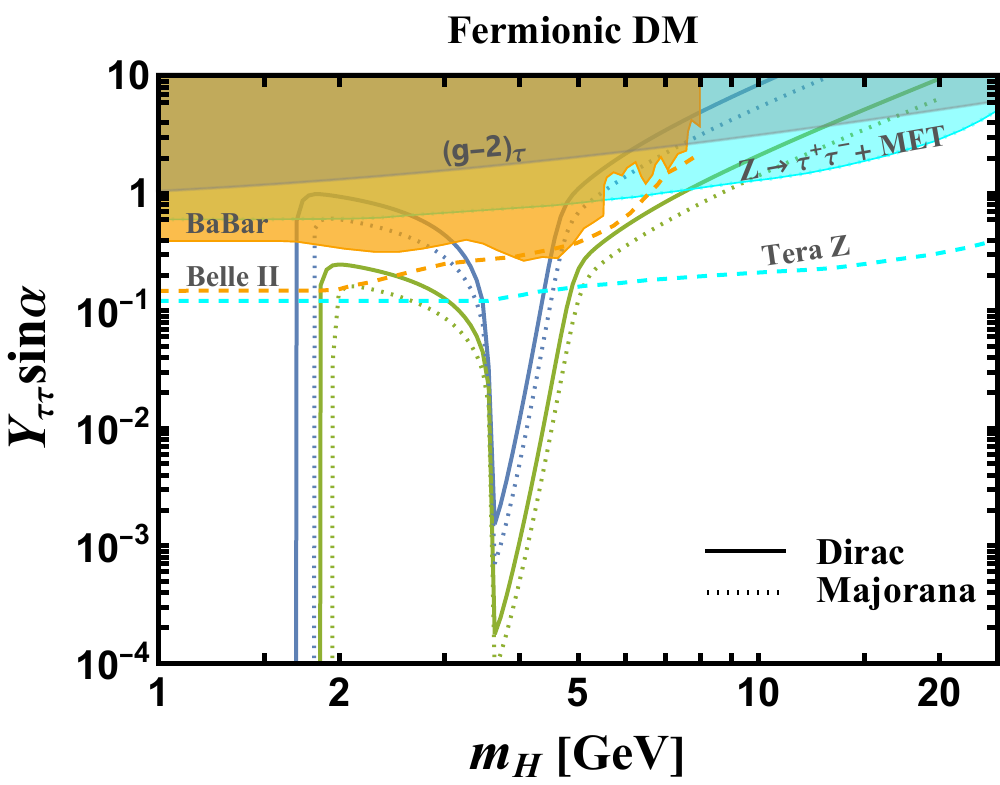}
$$
\caption{Theoretical and experimental constraints on the scalar (top) and fermionic forbidden DM (bottom)  for DM annihilations into $\mu^+\mu^-$ (left), $\mu +\tau$ (center), and $\tau^+\tau^-$ (right). 
The different colored contours represent the DM relic density $\Omega h^2 =0.12$ for different values of relative mass spliting: $\Delta=0.1$ (blue), and $\Delta=0.001$ (green).
The green band indicates the $(g-2)_\mu$ $2\sigma$ preferred range.
The other shaded regions regions are exclusion bounds: 
supernovae cooling~\cite{Kamiokande-II:1987idp,Croon:2020lrf} (purple), SLAC beam dump E137~\cite{Bjorken:1988as,Batell:2017kty} (brown), $(g-2)_{\mu}$ at Fermilab~\cite{Muong-2:2021ojo} (gray),$(g-2)_{\tau}$ ~\cite{Beresford:2019gww} (gray), $\tau$ decay width~\cite{Tanabashi:2018oca} (light pink), $\tau \rightarrow \mu H$ at ARGUS~\cite{ARGUS:1995bjh} (orange), searches for $Z\rightarrow \bar{\tau}\tau+\text{MET}$ at LEP~\cite{Tanabashi:2018oca,Chen:2018vkr} (cyan), dark photon searches through $e^+e^- \rightarrow \mu^+\mu^- H$ and  $e^+e^- \rightarrow \gamma H$ at \textsc{BaBar}~\cite{TheBABAR:2016rlg,BaBar:2017tiz} (light and dark yellow respectively).
We also show the constraints from \textsc{Planck}~\cite{Planck:2018vyg,Slatyer:2015jla} on radiative DM annihilation into two photons (blue/green regions), as well as the prospective reach of proposed MeV gamma-ray telescopes (blue/green dashed). For the scalar (fermion) DM scenario, we set $\kappa_{ij}=10^{-3}$ ($Y_{\chi}\cos{\alpha}=0.1$). 
} \label{relic-scalar-y}
\end{figure*}

\textbf{\emph{Results and phenomenological implications}.--}
Our model incorporates new physics below the electroweak scale, which has a myriad of phenomenological implications and makes it very predictive. This section analyzes and discusses the major results and the implications.  
Figure~\ref{relic-scalar-y} shows relic abundance predictions, contrasted with constraints from laboratory experiments and late time DM annihilation.
The panels show the values of the Yukawa couplings $Y_{\mu\mu,\mu\tau,\tau\tau}$ required to reproduce the observed relic abundance in the three leptophilic forbidden DM regimes $m_\mathrm{DM} \lesssim \{m_\mu,(m_\mu+m_\tau)/2,m_\tau\}$ (left to right) for both the scalar (top row) and fermionic (bottom row) DM scenarios and different mass splittings, as function of the mediator mass.
The $\Delta = 10^{-3} \sim 0$ line (green) indicates the lowest achievable couplings in the forbidden DM scenario. Larger $\Delta = 0.1$ (blue) results in a larger Boltzmann suppression, implying larger couplings. The smallest couplings $y\sim 10^{-5}$ are reached close to the $s$-channel resonance for $m_H \gtrsim 2 m_\mathrm{DM}$.
Towards low $m_H$, `secluded' annihilation~\cite{Pospelov:2007mp} into mediator pairs determines the relic abundance, which becomes independent of the coupling to leptons \footnote{The secluded parameter space is ruled out by indirect detection constraints on $S S \to H H \to 4\gamma$ in the case of scalar DM annihilating to $\tau\tau$ (drawn with solid gray vertical lines in the plot). In the other cases shown, the correct relic abundance is reached in the forbidden secluded regime.}.

DM indirect detection constraints on tree-level DM annihilation are avoided in the forbidden DM scenario. We point out that loop-level annihilation into two photons is a generic probe of forbidden DM coupled to SM fermions. \textsc{Planck} bounds~\cite{Planck:2018vyg,Slatyer:2015jla} on $\sigmav_{SS \to \gamma\gamma}$(see eqn.~\ref{eqn:sigmavgg} in the Appendix) during CMB decoupling are shown in Fig.~\ref{relic-scalar-y} as blue and green shaded regions. Galactic gamma ray lines at energies just below the lepton masses are a specific prediction of the present scenario, which could be probed at future MeV gamma ray missions like the AMEGO and e-ASTROGAM proposals~\cite{AMEGO:2019gny,e-ASTROGAM:2017pxr,Bartels:2017dpb} (dashed lines). In the fermionic DM scenario, annihilation is velocity suppressed, rendering this detection channel ineffective.

Due to the requirement of a light mediator, there are non-trivial consequences on the scalar mass-spectrum; moreover, to be consistent with all experimental constraints, we consider the possibility of leptophilic 2HDM, for which we assume the second Higgs doublet to have zero/negligible coupling with the quarks. 
A lower bound on the mass of the CP-odd scalar $A$ is obtained from $Z$ decay width measurements, $m_A \geq m_{Z}-m_H\simeq 90$ GeV \cite{Electroweak:2003ram,Tanabashi:2018oca}, where $m_Z$ is mass of the $Z$ boson. In addition, the LEP experiment set a lower bound on the charged scalar mass, $m_{H^+}\simeq 100$ GeV~\cite{Babu:2019mfe}. Additionally, there will be bounds from slepton searches at the LHC \cite{CMS:2018yan, ATLAS:2014hep,CMS:2018eqb}, which are less stringent for this sort of light leptophilic charged scalars (For details, see discussion in Ref.~\cite{Babu:2019mfe}). These lead to the mass hierarchy $m_{H}\ll m_{A},m_{H^+}$. A mass splitting of this type is  strongly constrained from the electroweak precision observables. The choice $m_{H}\ll m_{A}\simeq m_{H^+}$ with $\mathcal{O}(100)$ GeV mass splitting between $m_H$ and $m_{H^+}$ is however allowed \cite{Jana:2020pxx} from the $T$ parameter constraints, presenting no obstacles to the role of $H$ as a light mediator. 
Due to the alignment limit (no mixing of the SM Higgs $h$ and CP-even scalar $H$) and the absence of massive cubic scalar couplings involving the SM Higgs, the predictions for Higgs observables do not deviate from the SM predictions and are hence unconstrained by the LHC Higgs searches. For the rest of our analysis, we set $m_{A},m_{H^+} \simeq $110 GeV and vary the light CP-even scalar mass from the $\mathcal{O}$(MeV) to the $\mathcal{O}$(GeV) range. This unrestricted parametric space of the 2HDM has received little attention in the literature.

The light leptophilic charged scalar $\mathcal O(100)$ GeV can result in large non-standard neutrino interactions \cite{Babu:2019mfe} and generate Glashow-like resonance features  in  the ultrahigh energy neutrino event spectrum of future neutrino telescopes~\cite{Babu:2019vff,Huang:2021mki,Babu:2022fje}. This form of hierarchical scalar mass spectrum predicts the novel same-sign di-lepton signature $pp \to H^\pm H^\pm jj \to l_\alpha^\pm l_\beta^\pm j j + {E\!\!\!\!/}_{T}$ through same-sign pair production of charged scalars at the LHC~\cite{Jana:2020pxx}. This can be a good test of our model, but the detailed exploration is beyond the scope of this study.

In the following, we discuss other phenomenological implications of a light scalar coupling to SM leptons and DM.
Constraints are shown as bordered shaded regions in Fig.~\ref{relic-scalar-y}.
Light scalars coupled to charged leptons contribute to the lepton anomalous magnetic moments. For  coupling to the muon or muon and tauon, the loop corrections mediated by the light CP-even scalar $H$ always contribute positively to the muon anomalous magnetic moment.
In Fig.~\ref{relic-scalar-y}, we show the parameter space consistent with the observed muon anomalous magnetic moment measurement at Fermilab~\cite{Muong-2:2021ojo} as green band.
The grey shaded regions above this band indicate the parameter space where the $(g-2)_{\mu}$ discrepancy is larger than $5\sigma$, or $(g-2)_{\tau}$ is larger than the $2 \sigma$ constraint determined by~\cite{Beresford:2019gww}.

New scalars lighter than a few hundred MeV can be produced in astrophysical settings such as supernovae. Their possible contribution to supernova cooling is constrained by the neutrino observation of SN1987A~\cite{Kamiokande-II:1987idp}. Recently Ref.~\cite{Croon:2020lrf} studied the impact of new light particles interacting with muons on SN1987A, which we recast to the purple-shaded region in Fig.~\ref{relic-scalar-y}.

At collider or beam dump experiments, the light scalar can be produced in association with the leptons it couples to.
The null results from the electron beam dump experiment E137~\cite{Bjorken:1988as} constrain the $H \mu^+  \mu^-$ coupling~\cite{Batell:2017kty} (brown in Fig.~\ref{relic-scalar-y}).
Dark photon searches at BaBar \cite{TheBABAR:2016rlg} provide a stringent constraint on a light scalar with coupling to the muons through the $e^+  e^- \rightarrow \mu^+  \mu^- H$ process \cite{Batell:2016ove,Batell:2017kty}, with $H\rightarrow \mu^+  \mu^- $. We recast this result for the case $ \text{BR}(H\rightarrow \mu^+  \mu^-)< 1 $ (light-yellow in Fig.~\ref{relic-scalar-y}).
The BaBar collaboration also searches for events with a high energy monophoton and large missing energy~\cite{BaBar:2017tiz}, which puts limits on the $H \tau^+  \tau^-$ coupling~\cite{DAgnolo:2020mpt,Chen:2018vkr,Dolan:2017osp} (yellow in Fig.~\ref{relic-scalar-y}).  The dashed yellow line below this region indicates the projected sensitivity on the same process from the Belle-II experiment~\cite{DAgnolo:2020mpt,Dolan:2017osp,Belle-II:2010dht}. 

The $Y_{\tau \tau}$ coupling is also constrained from the $Z$ decay width measurement~\cite{Chen:2018vkr,DAgnolo:2020mpt}, where the associated production and subsequent dark decay of $H$ can contribute to the measured $Z\rightarrow \tau^+  \tau^-$ width~\cite{Tanabashi:2018oca} (cyan in Fig.~\ref{relic-scalar-y}). 
Finally, $\tau$ decays are modified in the $\mu\tau$-coupled case.
The search for lepton flavour violating two-body $\tau$ decay at ARGUS~\cite{ARGUS:1995bjh} requires $m_H > m_\tau - m_\mu$ to forbid $\tau\to\mu H$ decay (orange in Fig.~\ref{relic-scalar-y}).
Similarly, the decay $\tau\to\mu S S$ would contribute to the measured width $\Gamma( \tau \rightarrow \mu \bar{\nu}_{\mu}\nu_{\tau})$~\cite{Tanabashi:2018oca}, requiring $m_{\rm DM}> (m_{\tau}-m_{\mu})/2$ or equivalently $\Delta<0.126$ (light pink in Fig.~\ref{relic-scalar-y}).

In Fig.~\ref{relic-scalar-y}, we also show the projected sensitivities from various experiments. Muon beam dump experiments can be used to probe for a light scalar that has couplings to muons and DM by performing a muon missing energy search. The corresponding projected sensitivity from the muon beam experiment NA64-$\mu$~\cite{DAgnolo:2020mpt,Chen:2018vkr,Gninenko:2014pea} is shown in Fig.~\ref{relic-scalar-y} as a brown-dashed line. Future $Z$ factories based on $e^+e^-$ colliders look for exotic decay modes of $Z$ bosons. The projected sensitivity from these experiments~\cite{DAgnolo:2020mpt,Liu:2017zdh,TLEPDesignStudyWorkingGroup:2013myl,dEnterria:2016fpc,dEnterria:2016sca} is shown in Fig.~\ref{relic-scalar-y} as a cyan-dashed line.  

Overall, the light 2HDM-mediated forbidden DM scenario shares many implications with previous phenomenological models~\cite{DAgnolo:2020mpt}, but makes further predictions for the scalar sector around the electroweak scale.
From the DM side, radiative annihilation is a very informative probe of forbidden DM annihilating into SM particles, which in particular could identify $\mu\mu$-coupled scalar DM in the $(g-2)_\mu$-favoured part of parameter space in the future.

\textbf{\emph{Neutrino mass}.--}
The simplest radiative neutrino mass model, namely, the Zee model \cite{Zee:1980ai,Zee:1985id}, utilizes two-Higgs-doublets. Hence neutrino masses can arise 
naturally  by embedding our setup within the Zee model. A singly charged scalar needs to be added to our setup to complete the loop diagram that provides neutrino mass. We consider such a possibility and show that the same Yukawa couplings responsible for providing the correct  DM relic abundance give rise to non-zero neutrino masses.

In addition to the Yukawa couplings given in Eq.~\eqref{Yukawa}, the Zee model contains one more term, which is 
\begin{align}
-\mathcal{L}_Y\supset f_{ij}L_i \epsilon L_j \eta^+ +h.c.,\label{singly}    
\end{align}
where $\epsilon$ is the Levi-Civita tensor and $\eta^+\sim (1,1,1)$ is a singly charged scalar. The Yukawa coupling $f_{ij}$ is antisymmetric in flavor indices. Eqs.~\eqref{Yukawa} and \eqref{singly}, together with the following cubic term in the scalar potential,
\begin{align}
-V\supset \mu H_1 \epsilon H_2 \eta^- +h.c.,    
\end{align}
lead to non-zero neutrino mass given by, 
\begin{align}
&M_\nu=a_0\left( f m_E Y_l - Y_l^T m_E f  \right),   
\\&
a_0=\frac{\sin 2\omega}{16\pi^2} \ln\left(\frac{m^2_{h^+}}{m^2_{H^+}}\right); \;\;
\sin 2\omega=\frac{\sqrt{2}v\mu}{m^2_{h^+}-m^2_{H^+}},
\end{align}
where, $m_E$ is the diagonal charged lepton mass matrix,  $\omega$ is the mixing angle between the singly charged scalars, and $h^+$ and $H^+$ represent the mass eigenstates \footnote{For small mixing angle, which is typically required to provide tiny masses to neutrinos,  to a good approximation, the $H^+$-state can be identified with the charged guy coming from 2HDM.}.

DM annihilation into SM particles occurs through the Yukawa coupling $Y_l$; hence its texture is highly restricted from DM phenomenology. Moreover,  lepton flavor violating (LFV) processes put stringent limits on its off-diagonal entries required to satisfy neutrino oscillation data.
In the $\mu\mu$-coupled forbidden DM scenario,
the mediator mass must be close to twice the muon mass. In this case, the ARGUS experiment rules out any sizable non-zero entry with the tau-lepton due to null observation of $\tau\to \ell H$ processes, where $\ell=e, \mu$.
To generate viable neutrino masses and mixings, we hence consider a Yukawa texture with zero $2\times 2$ block in the 1-2 sector. 

Now, to be consistent with DM phenomenology, we fix $y_{\mu\tau}=5\times 10^{-4}$. As can be seen from Fig.~\ref{relic-scalar-y} (upper-middle plot), a coupling of this order correctly  reproduces the DM relic abundance with the mediator mass close to 2 GeV. Since the upper $2\times 2$-block is set to zero, reproducing the neutrino oscillation data requires the rest of the entries to be non-zero. However, each non-vanishing entry causes LFV  and is required to be small. Since the charged scalars are heavier than 110 GeV, LFV is entirely dominated by the light scalar $H$ in this theory. 
The most dangerous LFV processes are light scalar mediated $\mu\to e\gamma, \tau\to e\gamma,$ and $\tau\to \mu\gamma$ at one-loop, which we compute following  \cite{Dutta:2020scq}.

Following the discussion above, we perform a fit to the neutrino sector, which is simultaneously consistent with the DM relic abundance and provide a benchmark,
\begin{align}
&Y_l=10^{-4}
\begin{pmatrix}
0  &0 &3.494\times 10^{-4}      \\
0  &0&  5\\
-10^{-3}& -0.382&  0.542       
\end{pmatrix},
\\
&a_0\cdot f=10^{-7}
\begin{pmatrix}
0   &       2.135&  0        \\
 -2.135&  0        &  2.266\\
  0        & -2.266&  0         
\end{pmatrix}.
\end{align}
Neutrino observables associated with this fit yield,
\begin{align*}
&\Delta m^2_{21}=7.486\times 10^{-5} eV^2,\;
\Delta m^2_{31}=2.511\times 10^{-3}  eV^2, \nonumber 
\\
&\theta_{12}=34.551^\circ,\;
\theta_{23}=47.830^\circ,\;
\theta_{13}=8.545^\circ.
\end{align*}
These predictions are in good agreement with the current neutrino oscillation data \cite{Esteban:2020cvm}. While the rates of $\tau\to e\gamma$ and $\tau\to \mu\gamma$ can be easily suppressed, the branching ratio of $\mu\to e\gamma$ is typically close to the current limit \footnote{Current experimental limits on these LFV processes are \cite{ParticleDataGroup:2020ssz}: 
\begin{align*}
&BR(\mu\to e\gamma) < 4.2\times 10^{-13},\\
&BR(\tau\to e\gamma) < 3.3\times 10^{-8},\\
&BR(\tau\to \mu\gamma) < 4.4\times 10^{-8}.
\end{align*}} or within one order of magnitude smaller; therefore, future experiments such as MEG-II collaboration \cite{MEG:2016leq} will be able to test this theory.

\textbf{\emph{Conclusions}.--}
This work proposes a minimal realization of light Dark Matter, enabled by a light scalar mediator that can arise in the 2HDM. We focus on forbidden DM annihilating to SM leptons, which predicts the DM mass to be close to the $\mu$ or $\tau$ masses.
Stringent CMB constraints on sub-GeV DM are avoided, while we have identified Galactic gamma ray lines from radiative annihiation as a specific probe.
A distinctive trademark is that all new physics states appear at or below the EW scale; in particular, a CP-odd and charged scalars are predicted to have masses of order 100 GeV.
Furthermore, a leptophilic-like 2HDM of this type can shed light on the $(g-2)_\mu$ anomaly, and
--when embedded within the Zee model-- the couplings that determine the relic abundance become intimately linked to neutrino oscillations.
This minimal kinematically forbidden scenario is very predictive and in particular testable at future beam-dump experiments, colliders and sub-GeV gamma-ray telescopes.
However, the light 2HDM portal to Dark Matter is rather general and provides a simple way of linking light dark sectors to the Standard Model that may well have wider application.

\begin{acknowledgments}
{\textbf {\textit {Acknowledgments.--}}} The work of VPK is in part supported by US Department of Energy Grant Number DE-SC 0016013. The work of S.S. has been supported by the Swiss National Science Foundation. We acknowledge the use of the
 FeynCalc package \cite{Shtabovenko:2016sxi}. 
\end{acknowledgments}
\bibliographystyle{utphys}
\bibliography{reference}

\begin{widetext}

\section*{Appendix-I}\label{App-01}
\textbf{ Scalar potential:} 
The most general scalar potential (2HDM $+$ real scalar singlet) in the Higgs basis is given by 
\begin{align}
&V(H_1,H_2,S)= \mu_{1}^2H_1^{\dagger}H_1+\mu_{2}^2H_2^{\dagger}H_2
-\{\mu_{12}^2H_1^{\dagger}H_2+{\rm h.c.}\} 
+\frac{\lambda_1}{2}(H_1^{\dagger}H_1)^2
+\frac{\lambda_2}{2}(H_2^{\dagger}H_2)^2
+\lambda_3(H_1^{\dagger}H_1)(H_2^{\dagger}H_2)
\nonumber\\ & \quad
+\lambda_4(H_1^{\dagger}H_2)(H_2^{\dagger}H_1)
+\left\{\frac{\lambda_5}{2}(H_1^{\dagger}H_2)^2+{\rm h.c.}\right\}
+\left\{
\big[\lambda_6(H_1^{\dagger}H_1)
+\lambda_7(H_2^{\dagger}H_2)\big]
H_1^{\dagger}H_2+{\rm h.c.}\right\} \nonumber\\ & \quad
+ \frac{\mu_S^2}{2} S^2 + \frac{\lambda_S}{4!}S^4 + \frac{\kappa_1}{2} S^2 (H_1^{\dagger}H_1) + \frac{\kappa_2}{2} S^2 (H_2^{\dagger}H_2)
+  \left\{\frac{\kappa_{12}}{2} S^2 (H_1^{\dagger}H_2)+h.c.\right\}.
\end{align}

\smallskip
\textbf{Cross section $SS\to \gamma\gamma$:}
The velocity averaged $s$-wave annihilation cross section of scalar DM into two photons is given by (using~\cite{Resnick:1973vg}):
\begin{equation}
    \langle\sigma v\rangle_{SS\to\gamma\gamma}^{s\mathrm{-wave}} = \frac{e^4 \kappa_{12}^2 v^2 m_S^2 Y_{ll}^2 \left|\mathcal{I}[(2m_S)/m_l]\right|^2}{64 \pi^5 m_l^2 ((4 m_S^2-m_H^2)^2-\Gamma_H^2 m_H^2)}\,,
    \label{eqn:sigmavgg}
\end{equation} 
where
\begin{equation}
    \mathcal{I}[r] = \int_0^1 \mathop{dx} \int_0^{1-x}\mathop{dy} \frac{1 - 4 x y}{1 - x y r^2}\,.
\end{equation}

\section*{Appendix-II}\label{App-02}

\textbf{ Fermionic DM scenario:}
Fermionic DM candidates can be either Dirac or Majorana in nature. In this work, we consider both these possibilities. As usual, the stability of the DM is ensured by a $\mathcal{Z}_2$ symmetry. Similar to the scalar DM setup discussed in the main text, the 2HDM-portal is needed to implement the forbidden fermionic DM scenario. Since a singlet fermion $\chi$ cannot directly couple to the 2HDM sector, the presence of a scalar singlet $S=v_{S}+\omega$ is required. Then the relevant Yukawa couplings are given by, 
\begin{align}
	-\mathcal{L}_Y &\supset \widetilde{Y}_l \bar{\psi}_L H_1 \psi_R +  Y_l \bar{\psi}_L H_2 \psi_R + b\; Y_{\chi} S \bar{\chi}\chi 
	+ b\; m_{\chi} \bar{\chi}\chi + h.c.,
\end{align}
with $b=1 (1/2)$ for  Dirac (Majorana) fermion. The mass of the DM candidate $\chi$ can be written as,
\begin{align}
    & m_{\rm DM}=m_{\chi}+ Y_{\chi}v_S.
\end{align}

Like in the previous case, it is convenient to rotate the doublet scalars to the Higgs basis Eq. \ref{para}, then the full scalar potential can be written as,
\begin{align}
V(H_1,H_2,S)&=\mu_{1}^2H_1^{\dagger}H_1+\mu_{2}^2H_2^{\dagger}H_2
-\{\mu_{12}^2H_1^{\dagger}H_2+{\rm h.c.}\} 
+\frac{\lambda_1}{2}(H_1^{\dagger}H_1)^2
+\frac{\lambda_2}{2}(H_2^{\dagger}H_2)^2
+\lambda_3(H_1^{\dagger}H_1)(H_2^{\dagger}H_2)
\nonumber\\ & \quad
+\lambda_4(H_1^{\dagger}H_2)(H_2^{\dagger}H_1)
+\left\{\frac{\lambda_5}{2}(H_1^{\dagger}H_2)^2+{\rm h.c.}\right\}
+\left\{
\big[\lambda_6(H_1^{\dagger}H_1)
+\lambda_7(H_2^{\dagger}H_2)\big]
H_1^{\dagger}H_2+{\rm h.c.}\right\} \nonumber\\ & \quad 
+\frac{\mu_S^2}{2} S^2 +\frac{\mu_{SS}}{3!}S^3 + \frac{\lambda_S}{4!}S^4 + \mu_{1S} S (H_1^{\dagger}H_1)+ \mu_{2S} S (H_2^{\dagger}H_2)+\left\{\mu_{12S} S (H_1^{\dagger}H_2)+h.c.\right\} \nonumber\\ &\quad 
+\frac{\kappa_1}{2} S^2 (H_1^{\dagger}H_1) + \frac{\kappa_2}{2} S^2 (H_2^{\dagger}H_2)+  \left\{\frac{\kappa_{12}}{2} S^2 (H_1^{\dagger}H_2)+h.c.\right\}. 
\end{align}
In the alignment limit,  $\phi_1^0\approx h$ is the SM Higgs, which decouples from the other two CP-even scalars $\phi_2^0$ and $\omega$.
The latter two mix with each other, and the corresponding mass eigenstates are denoted as $H$ (with mass $m^2_{H}$) and $H'$ (with mass $m^2_{H'}$) and are given by,  
\begin{align}
    & H=\cos{\alpha}\; \omega + \sin{\alpha}\; \phi_2^0,\nonumber \\
    &  H'= -\sin{\alpha}\; \omega + \cos{\alpha}\; \phi_2^0,
\label{mixing}    
\end{align}
where the mixing angle $\alpha$ is defined as 
\begin{align}
&\sin 2\alpha= \frac{2v(\mu_{12S} +k_{12} v_S)}{ m_{H}^2-m_{H'}^2}. 
\end{align}
This mixing is essential for DM phenomenology since it allows the DM to annihilate into SM leptons. 

\end{widetext}

\end{document}